\documentclass[twocolumn,showpacs,preprintnumbers,amsmath,amssymb]{revtex4}
\usepackage{graphicx}% Include figure files
\usepackage{dcolumn}% Align table columns on decimal point
\usepackage{bm}% bold math

\begin{document}

\title{Primary Energy Spectra and Elemental Composition.\\GAMMA Experiment}
\author{S.V.~Ter-Antonyan}
 \email{samvel@yerphi.am}
 \homepage{http://gamma.yerphi.am/samvel/}
\author{R.M.~Martirosov, A.P.~Garyaka, V.~Eganov}
 \affiliation{Yerevan Physics Institute, 2 Alikhanyan Br. Str., 375036 Yerevan, Armenia}
\author{N.~Nikolskaya, T.~Episkoposyan}%
\affiliation{Moscow Lebedev Physics Institute, Russia}
\author{J.~Procureur}
\affiliation{Centre d Etudes Nucleaires de 
Bordeaux-Gradignan, Gradignan, France}
\author{Y.~Gallant}
\affiliation{Laboratoire de Physique Th$\acute{e}$orique et Astroparticules,
            Universit$\acute{e}$ Montpellier II, France}
\author{L.~Jones}
\affiliation{University of Michigan, Dept. of Physics, USA}

\date{\today}% It is always \today, today,
             %  but any date may be explicitly specified

\begin{abstract}
On the basis of the Extensive Air Shower (EAS) data observed by the
GAMMA
experiment the energy spectra and elemental composition of the
primary cosmic rays have been derived in the $10^3\div10^5$ TeV energy
range. Reconstruction of the primary energy spectra is carried out
in the framework of the SIBYLL and QGSJET interaction models and the
hypothesis
of the power-law steepening primary energy spectra. The obtained energy
spectra 
of primary $H, He, O, Fe$ nuclei along with the SIBYLL interaction model
agree with the 
corresponding extrapolations of known balloon and satellite data  
at the $\sim10^3$ TeV energies. 
The energy spectra obtained from the QGSJET model, show predominant proton 
composition of cosmic rays in the knee region. The evident rigidity-dependent
behavior of the primary energy spectra for both interaction models 
are displayed at the following rigidities: 
$E_R\simeq2400\div3000$ TV (SIBYLL) and $E_R\simeq3400\pm200$ TV (QGSJET).\\
Using parametric event-by-event method of the primary energy evaluation by  
measured $N_{ch},N_{\mu}(E_{\mu}>5 GeV, R<50m)$ and age ($s$) shower 
parameters, the all-particle energy spectra were obtained.\\
All presented results are derived taking into account the detector response, 
reconstruction uncertainties of EAS parameters and fluctuation of EAS development.  
\end{abstract}

% PACS, the Physics and Astronomy
                                 % Classification Scheme.
                              %Use showkeys class option if keyword
                              %display desired
\maketitle
%==================================================
\section{Introduction}
The investigation of the energy spectra and elemental composition
of primary cosmic rays in the knee region ($10^3\div10^5$ TeV)
remains to be one of the intriguing  problems of 
the modern high energy cosmic-ray physics. 
Despite the fact that these investigations
have been carried out for more 
than half a century, the data on the
elemental primary energy spectra at energies of $E>10^3$ TeV 
need improvement.
However, a bend of the all-particle
energy spectra at around $3\cdot10^3$ TeV (called the "knee")
at overall spectrum $\sim E^{-2.7}$ until the knee and $\sim E^{-3.1}$
beyond the knee, may be considered as an avowed fact. Moreover, 
assuming that the supernova explosion is the main source of the cosmic rays,
different theoretical models 
of the high energy cosmic-ray origin and 
propagation through the Galaxy, predict the rigidity-dependent 
steepening primary energy spectra in the knee region
\cite{Peters,Biermann1,Stanev}.\\
High statistical accuracies of the last EAS experiments 
\cite{KASCADE1,EAS-TOP,CASA}
already allowed us to infer that the rigidity-dependent
steepening energy spectra of primary nuclei can approximately
describe the observed EAS size spectra in the knee region
in the framework of conventional interaction models. However, the
accuracies
of the obtained elemental primary energy spectra are still
insufficient due to both the uncertainty of interaction model
and the accuracy of the solutions of the EAS inverse problem.\\
The Gamma facility (Fig.~1) was designed at the beginning of 90's in
the 
framework of the ANI experiment \cite{ANI} and the preliminary results
of EAS investigations presented in
\cite{GAMMA00,GAMMA01,GAMMA02,GAMMA03}.
The main characteristic features of the GAMMA experiment are the 
mountain disposition, symmetric location of the EAS detectors
and underground muon scintillation carpet that detects EAS 
muon components at $E_{\mu}>5$ GeV energies.\\
Here, the description of GAMMA facility, EAS inverse approach  
determining the primary energy spectra using the observed EAS data, and 
main results of investigation during 2002-2004 \cite{GAMMA02,GAMMA03}
are presented 
in comparison with the corresponding MC-simulated data in the 
framework of the SIBYLL \cite{SIBYLL} and QGSJET \cite{QGSJET}
interaction models. 
%==================================================
\section{GAMMA experiment}
The GAMMA installation is a ground based array of 33 surface 
particle detection stations and 150 underground muon detectors 
located on the south side of Mount Aragats, Armenia.
Elevation of the GAMMA facility is 3200 m above sea level,
which corresponds to 700 g/cm$^2$ of atmospheric depth. The 
diagrammatic layout is shown in Fig.~1. \\
\begin{figure*}
%\begin{center}
%\epsfig{file=fig1.eps,width=14.714cm,height=6cm}
\includegraphics[width=14.714cm,height=6cm]{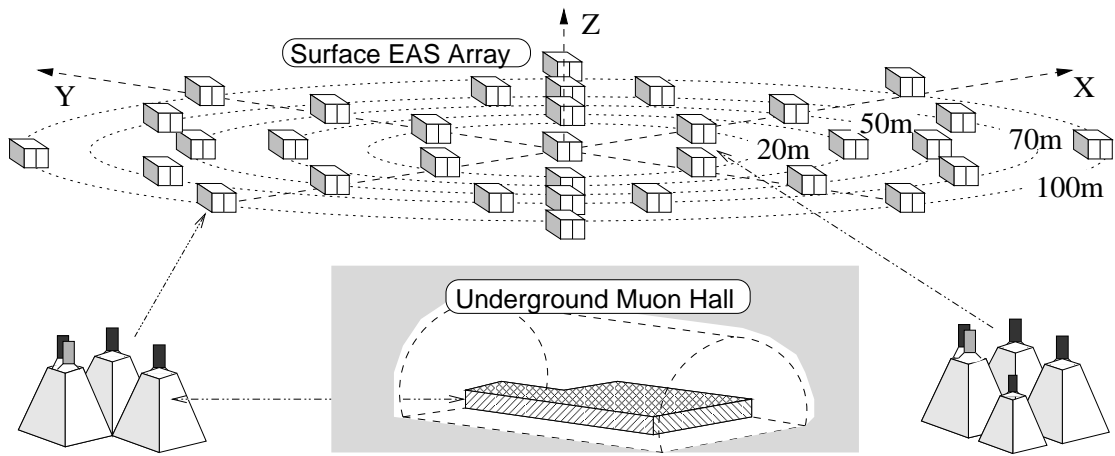}
%\end{center}
\caption{Diagrammatic layout of the GAMMA facility}
\end{figure*}
The surface stations of the EAS array are located on 5
concentric
circles of radii: 20, 28, 50, 70, 100 m and each station contains 
3 square plastic scintillation detectors with the following dimensions:
1x1x0.05 m$^3$. Each of the central 9 stations contains an additional
(4-th) small scintillator with dimensions 0.3x0.3x0.05 m$^3$ (Fig.~1) for high
particle density ($\gg10^2$ particles/m$^2$) measurements.\\
The photomultiplier tube is positioned on the top of the aluminum
casing covering the scintillator. One of the three station's detectors
is examined by two photomultipliers, one of which is designed 
for fast-timing measurements.\\
150 underground muon detectors (muon carpet) are
compactly arranged in the underground
hall under 2.3 Kg/cm$^2$ of concrete and rock. The dimensions, casing and 
applied photomultipliers are the same as in the EAS
surface detectors.
\subsection{Detector system and triggering}
The output voltage of each photomultiplier is converted into 
the pulse burst by logarithmic ADC and transmitted to the CAMAC array
where
the corresponding electronic counters produce a digital number
("code") of pulses in the burst. Four inner ("trigger") stations are 
monitored by a coincidence circuit. 
If each of at least two scintillators of each  
trigger station detects more than 3 particles, the information from 
all detectors are then recorded along with the time between the 
master trigger pulse and the pulses from all fast-timing detectors. 
The given trigger condition provides EAS detection with the EAS size
threshold $N_{ch}>(0.5\div1)\cdot10^5$ at the location of the EAS core
within the $R<25$ m circle. \\
Before being placed on the scintillation 
casing, all photomultipliers are tested by a test bench using a luminodiode method where 
the corresponding parameters of logarithmic ADC and the upper limits 
($(0.5\div1)\cdot10^4$) of the measurement ranges are determined. 
The number of charged particles ($n_i$) passing through the i-th 
scintillator is computed using a logarithmic 
transformation: $\ln n_i=(C-C_0)/d$, where 
the scale parameter $d\simeq(9\div10)\pm0.35$ is preliminarily
determined
by the test bench, $C=(0\div2^7-1)$ is an output digital 
code from the CAMAC array corresponding 
to the energy deposit of $n$ charged particles into the 
scintillator, $C_0\simeq(5\div6)\pm0.25$ is determined 
for each hour of run and is equal to 
the mode of the background single particle digital code spectra
(Fig.~2).\\
The time delay $\Delta t_j=t_j-t_1$ of each $j$-th ($j=2,\dots,33$) 
fast timing detector is estimated by the pair-delay method 
\cite{Ter-Antonyan1} at the resolution time about $4\div5$ ns.
\subsection{Reconstruction of EAS parameters}
EAS zenith angle ($\theta$) is estimated
on the basis of measured shower front arrival times
by 33 fast-timing surface detectors,
applying the maximum likelihood method and flat-front approach
\cite{Ter-Antonyan1,MAKET}.
The corresponding uncertainty are tested by MC simulations and is equal
to: $\sigma(\theta)\simeq1.5^0$.\\
The reconstruction of the EAS size ($N_{ch}$), shower age ($s$) and
core coordinates ($x_0,y_0$) are performed based on the 
NKG approximation of measured charged particle densities
($\{n_i\},i=1,\dots,m$) using the $\chi^2$ minimization 
to estimate $x_0,y_0$ and the maximum
likelihood method to estimate the $N_{ch}$ taking
into account the measurement errors. 
The  logarithmic transformation $L(n_i)=\ln n_i-(1/m)\sum\ln
n_i$ at $n_i\neq0$, allows to obtain the analytical solution
for the EAS age parameter ($s$) at the $\chi^2$ minimization
\cite{MAKET,Ter-Antonyan2}.
Unbiased ($<5\%$) estimations of $N_{ch},s,x_0,y_0$ shower
parameters are obtained at $N_{ch}>5\cdot10^5$, 
$0.3<s<1.6$, and $R<25$ m from the shower core to the 
center of the EAS array distances. 
Corresponding accuracies are derived from MC simulations by
the CORSIKA(EGS) \cite{CORSIKA} and are equal to:
$\Delta N_{ch}/N_{ch}\simeq0.1$, 
$\Delta s\simeq0.05$, 
$\Delta x,\Delta y\simeq0.5\div1$ m.\\
The reconstruction of the total number of EAS muons ($N_{\mu}$) by 
the detected muon densities ($\{n_{\mu,j}\}, j=1,\dots,150$)
from the underground muon hall is carried out by restricting
the distance $R_{\mu}<50$ m from the shower core
(so called "truncated" EAS muon size \cite{KASCADE01}) and 
Greisen approximation of the muon lateral distribution function: 
$\rho_{\mu}(r)=cN_{\mu}(R<50m)\exp{(-r/r_0)}/(r/r_0)^{0.7}$, where
$r_0=80$ m, $c=1/2\pi\int_{0}^{50} \rho(r)rdr$. 
The truncated muon size $N_{\mu}(R<50m)$ is estimated
at known (from the EAS surface array) shower core coordinates in the
underground muon hall. The unbiased estimations for muon size 
are obtained at $N_{\mu}>10^3$ using the maximum likelihood method 
and assuming Poisson fluctuations of detected muon numbers. 
The reconstruction accuracies of the truncated
muon size are equal to $\Delta N_{\mu}/N_{\mu}\simeq0.2\div0.35$
at $N_{\mu}\simeq10^5\div10^3$ respectively.\\
Notice, that the detected muons in the 
underground hall are always accompanied by the electron-positron 
equilibrium spectrum which is produced when muons pass through the 
matter (2300 g/cm$^2$) over the scintillation carpet. Since
this spectrum depends on the muon energy ($\sim\ln{E_{\mu}}$),
overestimations ($\sim 25\div30\%$) of the reconstructed muon size
have to depend on the primary energy and therefore on the EAS size. 
%==================================================
\section{EAS simulation}
\subsection{Key assumptions} All observed quantities
($\Delta F/\Delta\tilde{q}_u$) 
in the high energy EAS physics are obtained via convolutions of the 
energy spectra $dI_A/dE$ of primary nuclei ($A\equiv H,He,\dots$ at
least up to $Ni$) with the differential spectra $W_A(E,q_u)$ of 
the EAS parameters $q_u\equiv N_{ch}, N_\mu, s$ at the observation level and
EAS array response functions $\partial {\cal 
R}_A(E,q_u,\theta)/\partial\tilde{q}_u$: \\
\begin{equation}
\frac{\Delta F_u}{\Delta\tilde{q}_u}=
\frac{1}{C}
\sum_{A}\int_{E} \frac{dI_A}{dE}
\int_{D}\int_{Q_u} W_A(E,q_u)
\frac{\partial {\cal R}_A}{\partial\tilde{q}_u}
dEdDdq_u\;,
\end{equation}
where the EAS parameter $\tilde{q}_u$ is a reconstructed value of 
the corresponding parameter $q_u$ on the observation level, 
$dD\equiv\cos{\theta}dxdyd\Omega$ is an element of the multidimensional phase 
space 
($D$) of the EAS detection taking into account the EAS selection
criteria and trigger conditions,  $W_A(E,q_u,\theta)$  
are the corresponding differential spectra of the EAS parameters ($q_u$) 
at the primary energy $E$, zenith angle of incidence $\theta$ and 
a given kind of primary nucleus ($A$),
C is a corresponding normalization factor. 
In the general case, $W_A$ depends on the interaction model 
\cite{Ralph,Ter-Antonyan3}.\\
The multidimensional integral above is better to calculate by
Monte-Carlo simulation, especially since the spectra 
$W_A(E,q_u,\theta)$ can be computed more or less precisely only by 
3-dimensional EAS simulations.
\subsection{Simulation scenario}
We have computed the shower spectra $dW_A(E,q_u,\theta)$, ($q_u\equiv N_{ch},
N_\mu,
s\dots$) on the observation level of the GAMMA facility using the 
CORSIKA6031(NKG,EGS) EAS simulation code \cite{CORSIKA} 
with the QGSJET01 \cite{QGSJET} and SIBYLL2.1 \cite{SIBYLL} 
interaction models for 4 groups ($A\equiv H,He,O,Fe$) of primary
nuclei at the power-law energy spectra  
($\sim E^{-1.5}$) in the $5\cdot10^2\div5\cdot10^5$ TeV energy range.
The spectral index (-1.5) was chosen to provide high statistical
accuracies of the simulated data beyond the knee.\\   
The EGS mode of the CORSIKA was used for computations of the
response functions of the GAMMA detectors taking into account the EAS
gamma-quanta contributions and the choice of the corresponding input parameters
of the adequate NKG mode. \\
Each EAS particle ($\gamma,e,\mu,h$) obtained from the CORSIKA(EGS) on 
the observation level (not interrupting the CORSIKA routine) 
is passing through the 
steel casing (1.5 mm) of detector station and then through the 
corresponding scintillator. The pair production and Compton scattering 
processes are additionally simulated in the case of the 
EAS $\gamma$-quanta passing through the steel casing and the scintillator.\\  
The resulting energy deposit in the scintillator is converted to the ADC 
code and inverse decoded into a number of "detected" charge particles 
taking into account all uncertainties of the ADC parameters ($C_0,d$) 
and the fluctuation of the light collected by 
the photomultiplier ($\sigma_l\simeq0.25/\sqrt{n}$).  \\
Using the simulation scenario above, 100 EAS events were simultaneously 
simulated by the CORSIKA routine at the  
EGS and NKG modes  for $A\equiv H,He,O,Fe$ primary nuclei at 
log-uniform energy spectra in the $5\cdot10^2\div10^5$ TeV 
energy range. The computations of the charged particle 
densities in the surface detectors at NKG mode of the CORSIKA were 
performed by applying the two-dimensional  
interpolations of the corresponding particle density matrix from the
CORSIKA routine \cite{CORSIKA}.\\ 
The agreement ($\sim5\% $) of the EGS and NKG simulated data   
was attained at the $E_e\simeq1\pm1$ MeV kinetic energy
threshold of the EAS electrons (positrons) at NKG mode (input parameter 
of CORSIKA code). 
However, the energy threshold for the detection of a vertical single  
minimal ionizing background particle by scintillation counters is about 
$8\div9$ MeV and differences obtained by the CORSIKA NKG  
prediction are completely explained by contribution of EAS
$\gamma$-quanta.\\ 
Thus, the EAS simulations by the CORSIKA with fast
computation NKG mode at $E_e>1$ MeV  
is adequate to the EAS simulation by the EGS mode taking into account
the EAS $\gamma$-quanta and peculiarity of the GAMMA surface array.\\        
All EAS muons with energies of $E_{\mu}>4$ GeV on the GAMMA observation
level have passed through the  
2.3 Kg/cm$^2$ of rock to the muon scintillation carpet (the underground 
muon hall). 
Fluctuations of the muon ionization losses and electron (positron)
accompaniment due to the muon bremsstrahlung,  
direct pair production, knock on and photo-nuclear interactions
are taken into account. 
The transformation of the energy deposit  
to the number of detected muons is performed the same way as 
for the surface detectors.\\
The EAS simulations were performed at  
$4.5\cdot10^4$ primary $H$, $4.3\cdot10^4$ $He$,
$2.4\cdot10^4$ $O$, $2.4\cdot10^4$ $Fe$ nuclei 
using the CORSIKA NKG routine at the SIBYLL interaction model.
Corresponding statistics at the QGSJET interaction model
were: $4.1\cdot10^4$, $4.2\cdot10^4$, $2\cdot10^4$, $2\cdot10^4$.
The energy thresholds of primary nuclei were the same for both
interaction models and were set 0.5, 0.7, 1, 1.2 PeV
respectively at $5\cdot10^3$ PeV upper energy limit.
%==================================================
\section{Measurement errors and density spectra}
The close disposition of $k=1,2,3$ scintillators in each of the ($i$-th) 
detector station of the GAMMA surface array  
allows to auto-calibrate the measurement error by detected EAS data. 
The measured and simulated particle
density divergences $(n_{k}-\rho)/\rho$ versus 
average value $\rho=(1/3)\sum{n_{k}}$ at
$R_i>10$ m distances from shower core are shown in Fig.~2 (circle symbols). 
The obtained dependences are completely determined 
by Poisson fluctuations (at $R_i\gg1$ m ) and measurement errors.\\
\begin{figure}[h]
%\begin{center}
\includegraphics[width=7cm,height=7cm]{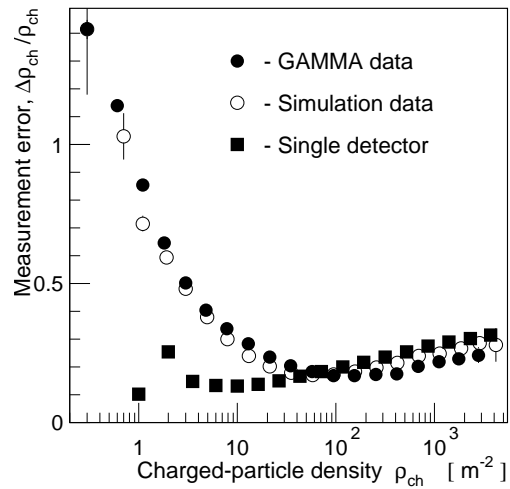}
%\end{center}
\caption{Particle density divergences 
(circle symbols) and measurement error of single detector (square symbols) versus 
charged-particle density.}
\end{figure}
The agreement of the measured and simulated dependences allowed to
extract the real measurement errors of the GAMMA 
detectors. In Fig.~2 the corresponding results are shown (square symbols).\\
The background single particle spectra (in the
units of ADC code) detected by GAMMA  
surface scintillators for 78 sec operation time are shown 
in Fig.~3 (dotted lines).
The background single particle spectra detected by underground 
\begin{figure}[h]
%\begin{center}
\includegraphics[width=7cm,height=7cm]{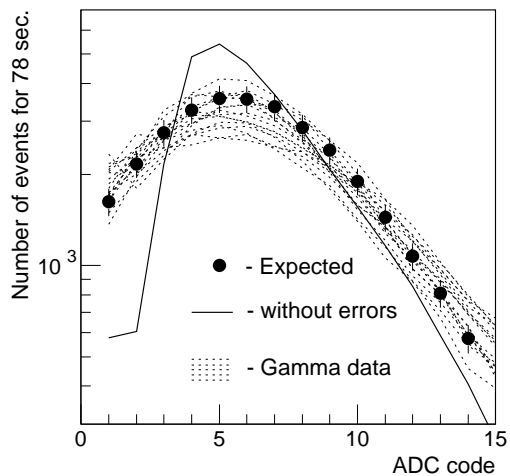}
%\end{center}
\caption{Background single particle spectra of 15 surface detectors
(dotted lines). The symbols (solid line) are the expected spectra 
taking into account (without) measurement errors.}
\end{figure}
muon scintillators have the same shape at about 10 times less
intensities.\\
These spectra are used for the operative (each hour) 
determination of ADC parameters ($C_0$) during an
experiment. The symbols and solid lines in 
Fig.~3 display the corresponding expected spectra 
obtained by MC-simulation taking into account the measurement errors
(symbols)
and without errors (line) respectively. The minimal primary energy in 
simulation of the background particle spectra was confined to the 7.6 
GV primary particle's geomagnetic rigidity.\\
Because the effective primary energies responsible for the 
single particle spectra at observation level 700 g/cm$^2$ are about 
$\sim100$ GeV and the energy range
is studied by direct measurements in the balloon and satellite
experiments, the primary energy spectra and 
elemental composition at MC-simulation were taken from approximations 
\cite{Wiebel}. Notice, that the expected single particle spectra
at these energies are practically the same for QGSJET and SIBYLL
interaction models.\\
Fig.~4a,b (symbols) display the charged particle density spectra
detected by the corresponding surface
detectors (a) and underground muon detectors (b) at $R_i<50$ m and 
different EAS size thresholds: 
$N_{ch}>5\cdot10^5$, $N_{ch}>10^7$ (and additionally 
$N_{ch}>2\cdot10^6$ for muon density spectra).\\
\begin{figure*}
%\begin{center}
\includegraphics[width=13cm,height=7cm]{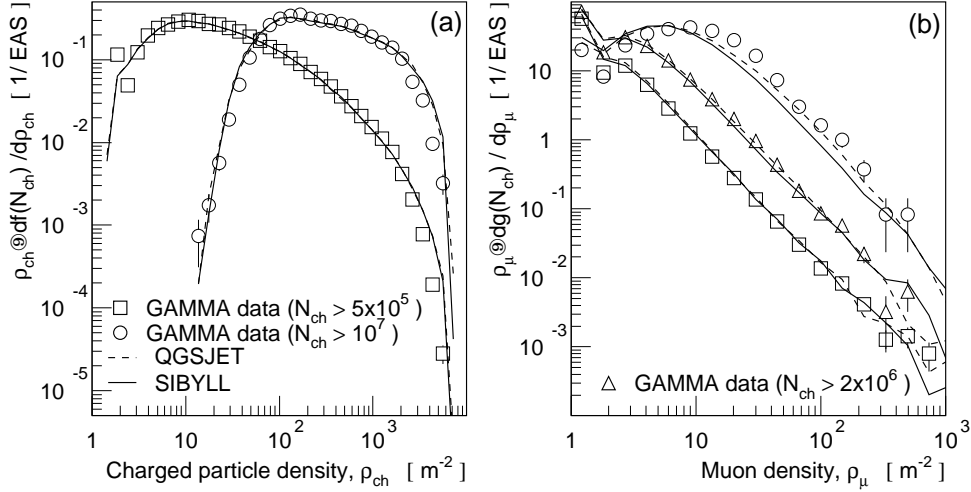}
%\end{center}
\caption{Detected (symbols) and expected (lines) 
particle density spectra of surface scintillators 
(left panel, a) and underground muon scintillators (right panel, b).}
\end{figure*}
The showers were selected at $\theta<30^0$ and the shower core
location in the $R<25$ m range from center of the GAMMA facility (Fig.~1). 
The corresponding expected spectra 
(lines) at different interaction
models are also shown in Fig.~4. The primary energy spectra
and 
elemental composition
at MC-simulations were taken from EAS inverse problem solution 
(see below). There is 
a good agreement of the expected and observed data for the surface
array in the measurement range (about four orders of
magnitude).  However, the agreement of the detected 
muon density spectra with expected ones 
is attained only in the $N_{ch}<10^7$ range. The 
observed discrepancies for the muon density spectra at $N_{ch}>10^7$ 
are unaccounted for the present and demand subsequent
investigations.
%==================================================
\section{EAS data }
The main EAS data of the GAMMA experiment 
are shown in Fig.~5-10 (symbols).
\begin{figure}
%\begin{center}
\includegraphics[width=7cm,height=7cm]{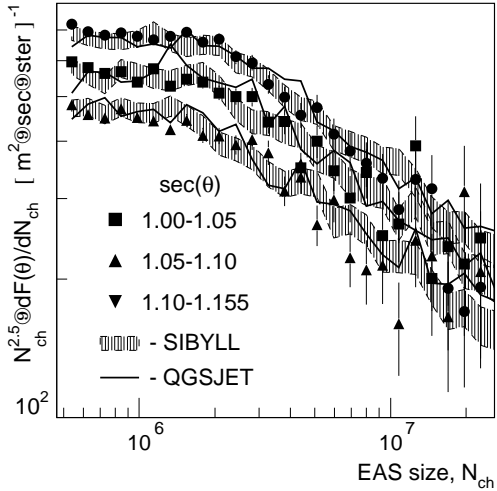}
%\end{center}
\caption{EAS size spectra at 3 zenith
angular intervals (symbols) and corresponding expected spectra
according to the SIBYLL (shaded area) and QGSJET interaction models 
(lines).} 
\end{figure}
These results were obtained at the $6.19\cdot10^7$ sec
operation time and following selection
criteria: $N_{ch}>5\cdot10^5$, $R<25$ m, 
$\theta<30^0$, $0.3<s<1.6$. All the lines and shaded areas in Fig.~5-10 
correspond
to the expected spectra according to the QGSJET and 
SIBYLL interaction models.\\
The EAS size spectra ($N^{2.5}_{ch}\cdot dF(\theta)/dN_{ch}$) at
3 zenith angular intervals are shown in Fig.~5.  
\begin{figure}
%\begin{center}
%\epsfig{file=fig6c.eps,width=6cm,height=6cm}
\includegraphics[width=7cm,height=7cm]{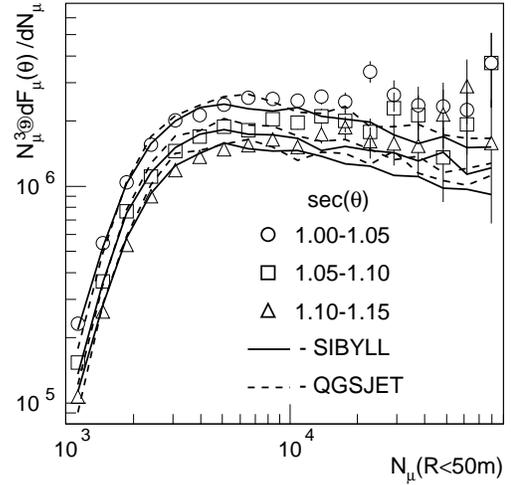}
%\end{center}
\caption{Normalized EAS truncated muon size spectra at 3 zenith
angular intervals (symbols). The lines correspond 
to the expected spectra at the SIBYLL
(solid) and QGSJET (dashed) interaction models.}
\end{figure}
The truncated muon size spectra  
in the same  zenith angular intervals are shown in Fig.~6. 
These spectra normalized to the EAS intensity at
$N_{ch}>5\cdot10^5$ and $\theta<30^0$.
The EAS size spectra at $\theta<30^0$ and 
different thresholds of the truncated EAS muon size are shown in 
Fig.~7. 
\begin{figure}
%\begin{center}
%\epsfig{file=fig7c.eps,width=6cm,height=6cm}
\includegraphics[width=7cm,height=7cm]{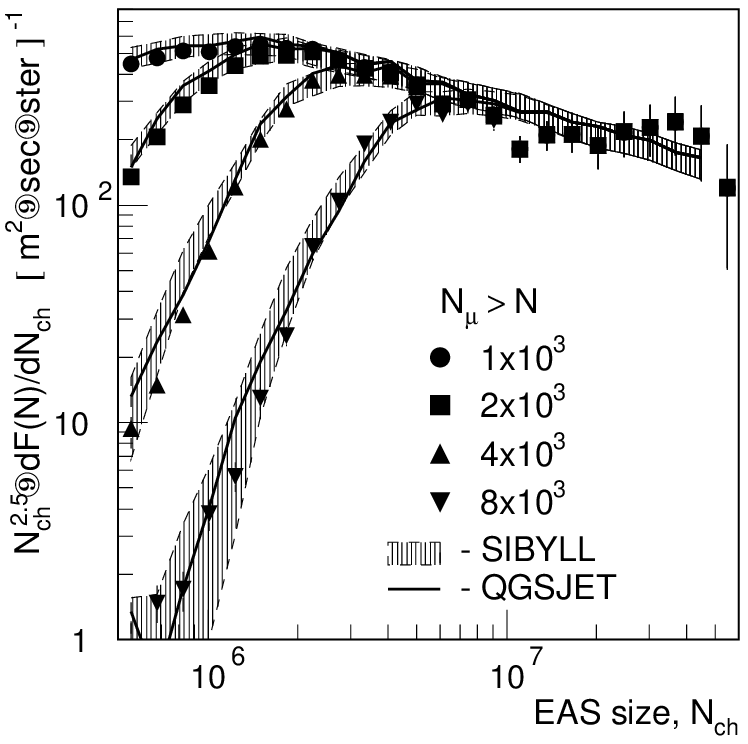}
%\end{center}
\caption{EAS size spectra (symbols)
at different truncated muon size
thresholds and $\theta<30^0$. The lines and shaded areas are the expected
spectra according to the QGSJET and SIBYLL interaction models.}
\end{figure}
The normalized EAS truncated muon size spectra
at different EAS size thresholds are shown in Fig.~8. 
\begin{figure}
%\begin{center}
%\epsfig{file=fig8c.eps,width=6cm,height=6cm}
\includegraphics[width=7cm,height=7cm]{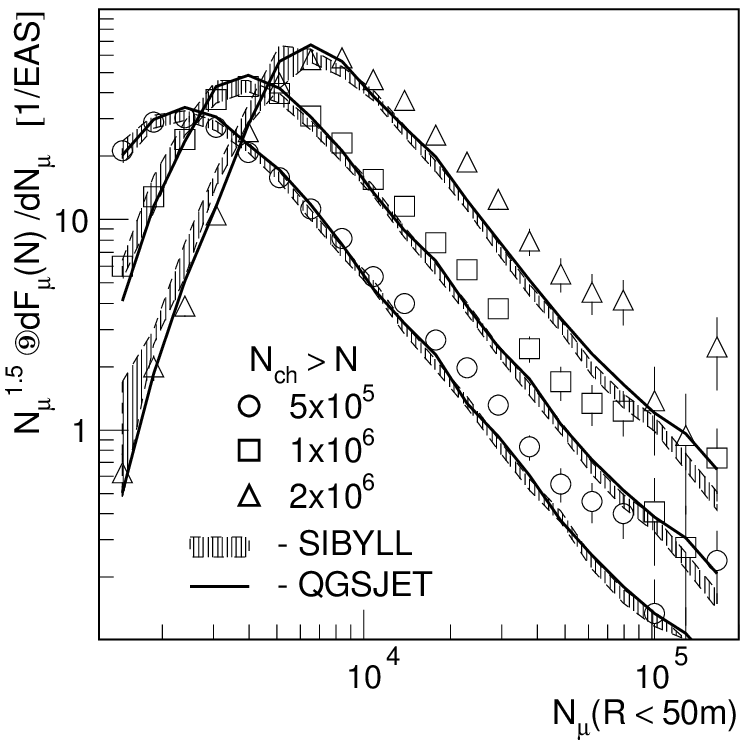}
%\end{center}
\caption{Normalized EAS truncated muon size spectra at different EAS
size thresholds (symbols). The lines and shaded areas correspond 
to the expected spectra at the QGSJET
and SIBYLL interaction models respectively.}
\end{figure}
Fig.~9 displays the average EAS age parameter dependence  
on EAS size. The lines are the expected dependences
according to QGSJET (dotted line) and SIBYLL (solid line) models.  
\begin{figure}
%\begin{center}
%\epsfig{file=fig9c.eps,width=6cm,height=6cm}
\includegraphics[width=7cm,height=7cm]{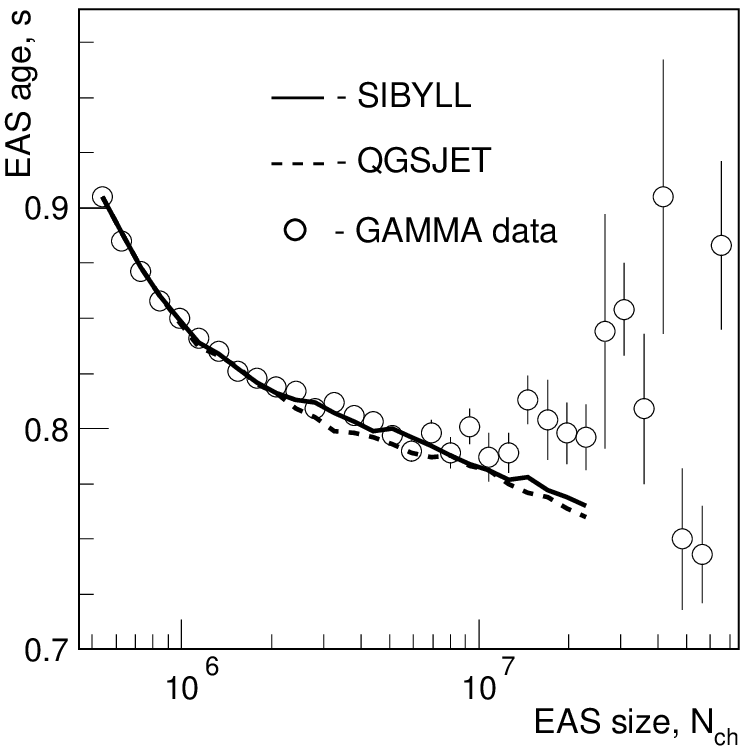}
%\end{center}
\caption{Dependence of average EAS age parameter on EAS size.}
\end{figure}
The obtained $N_{\mu}(N_{ch})$ dependences and corresponding
expected values at the primary
Hydrogen, Iron and mixed compositions computed in the frame of
the SIBYLL and QGSJET interaction models are plotted on
Fig.~10.
\begin{figure}
%\begin{center}
%\epsfig{file=fig10c.eps,width=6cm,height=6cm}
\includegraphics[width=7cm,height=7cm]{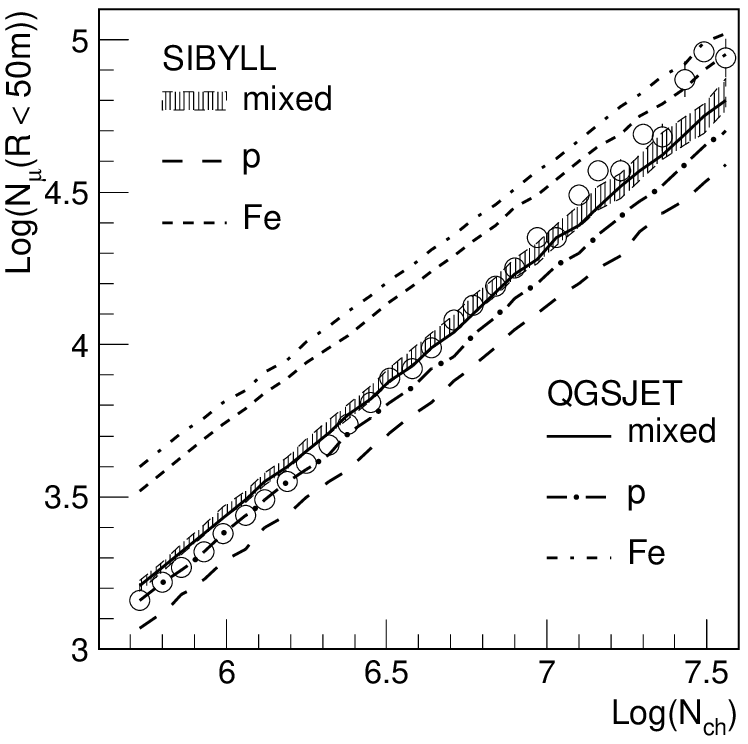}
%\end{center}
\caption{Average EAS truncated muon size 
$N_{\mu}(E>5$ GeV, $R<50$m) versus EAS size ($N_{ch})$. The dashed lines
and shaded area are the expected dependences at the SIBYLL interaction 
models and pure $p$, $Fe$ and mixed compositions respectively. Dash-dotted
and solid lines are the same dependences at the QGSJET models.}
\end{figure}
%================================================
\section{EAS inverse problem and primary energy spectra}
\subsection{Combined approximations of EAS data}
Direct computations of the expected EAS spectra 
using the integral expression (1)
is possible only in the framework of a given interaction model and
known primary energy spectra.
Moreover, the Gamma data shown in Fig.~4-10 may only formally compare 
with the same data obtained by other EAS experiments performed at both 
similar atmospheric depths \cite{Norikura,Tien-Shan,
MAKET} and depths close to the sea level \cite{KASCADE1, 
EAS-TOP,AKENO}. The
correct comparison is possible only at known primary energy
spectra and known interaction model because both transformation of
the detected EAS spectra to the spectra at a given observation level
and the extrapolation of the obtained spectra to an another atmosphere
depth in a general case are folded by the integral expressions similar
to (1).\\
In such case the more reliable way to interpret the
experimental data is to unfold the integral expression (1)
at a given interaction model. 
As a criterion of the validity of the solutions, the $\chi^2$ test 
of the detected and expected data may be performed. The agreement
between the obtained energy spectra at different primary nuclei and 
the corresponding extrapolations of known balloon and satellite data to the
given measurement range  will also validate the solutions.\\
Evidently, the accuracies of the unfolding
of expression (1) depend not only on number of measurement points (bins)
and different measured spectra but also on the wealth of
information about the primary energy spectra and the interaction model
involved in the given measured EAS spectra. The amount of information 
contained in the expression (1) reveals itself via stability
and uncertainties of the solutions.\\
\begin{figure*}
%\begin{center}
%\epsfig{file=fig11c.eps,width=12cm,height=7cm}
\includegraphics[width=12cm,height=7cm]{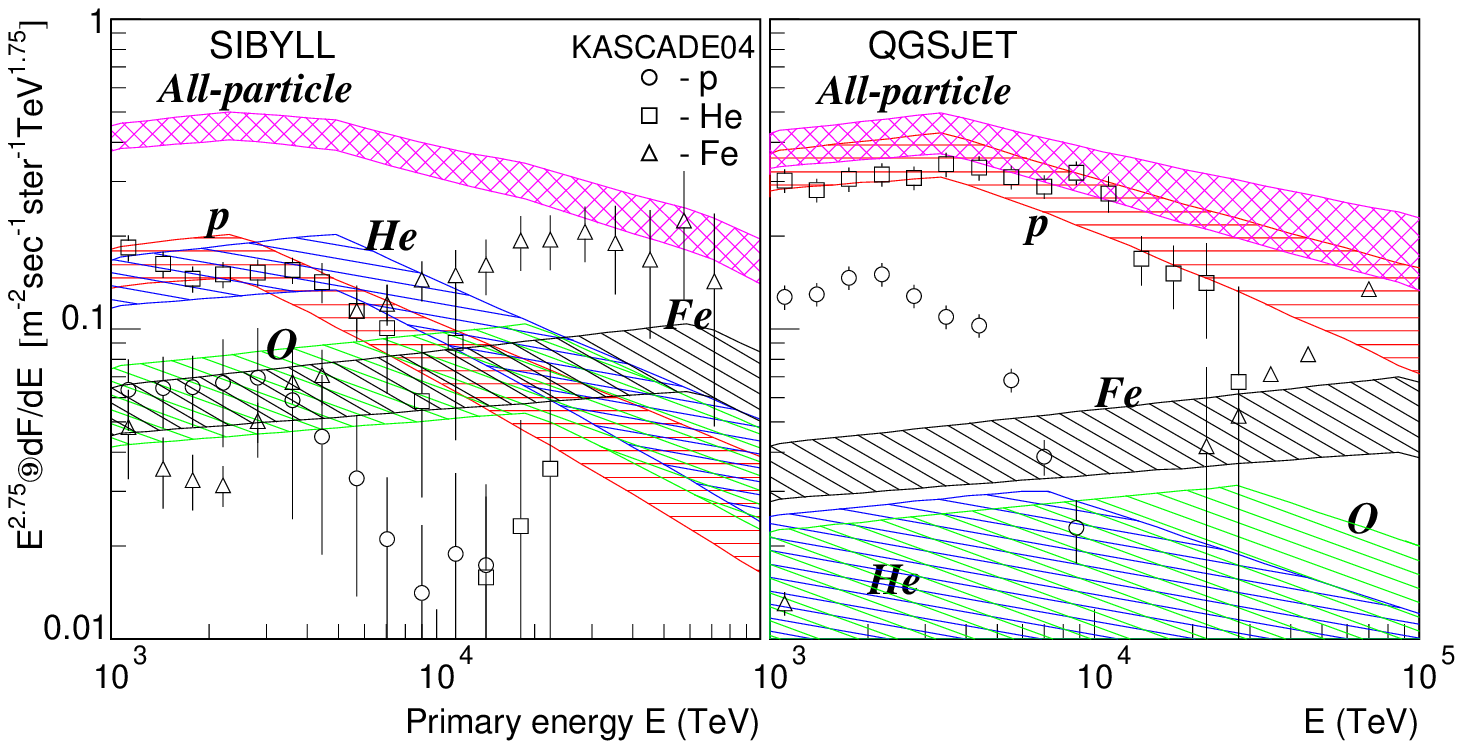}
%\end{center}
\caption{Energy spectra and abundance of the primary
nuclei (shaded areas) at the SIBYLL (left panel) and
QGSJET (right panel) interaction models. The symbols are the KASCADE
data from \cite{KASCADE04}.}
\end{figure*}
It is shown in \cite{Ter-Antonyan4}, that the EAS size spectra and EAS
truncated muon size spectra at three zenith angular intervals allow
to reliably unfold expression (1) at a given interaction model
for not more than 2 kinds of primary nuclei. The unreliability
of solutions of (1) for 4 kinds of primary nuclei was shown in \cite{Schatz}, 
as well.\\ 
Taking into account the above, we used the parameterization of the integral 
equation (1)
similar to \cite{Ter-Antonyan3, Ter-Antonyan5}. The solutions
for the primary energy spectra in (1) were sought based on
the theoretically known power-law function \cite{Biermann1} with
the "knee" at the rigidity-dependent energies $E_k(A)=E_R\cdot Z$ and the same 
indices ($-\gamma_1$) and ($-\gamma_2$) before and after the knee 
respectively, for all kinds of primary nuclei ($A$):
\begin{equation}
\frac{dI_A}{dE}=\Phi_AE_k^{-\gamma_1}\Big(\frac{E}{E_k}\Big)^{-\gamma}
\end{equation}
where $\gamma=\gamma_1\equiv2.65$ at $E\leq E_k(A)$,
$\gamma=\gamma_2$ at $E>E_k(A)$, $E_R$ is particle's rigidity
and $Z$ is a charge of A nucleus.\\
Thus, the integral equation (1) is transformed into a parametric
equation with unknown spectral parameters: $\Phi(A),E_k(A),\gamma_2$,
which are determined by minimization of $\chi^2$ function:
\begin{equation}
\chi^2=\frac{1}{\sum{V_u}}\sum_{1}^{U}\sum_{1}^{V_u}
\frac{(\zeta_{u,v}-\xi_{u,v})^2}
     {\sigma^2(\zeta_{u,v}) +\sigma^2(\xi_{u,v})}
\end{equation}
where $U$ is  the number of examined functions 
$\zeta_{u,v}\equiv\Delta F_u/\Delta\tilde{q}_{u,v}$ (Fig.~5-10, symbols) 
obtained from the experiment with statistical accuracies 
$\sigma(\zeta_{u,v})$ at $v=1,\dots,V_u$ measured points (bins), and
$\xi_{u,v}$ and $\sigma(\xi_{u,v})$ are the corresponding 
expected values of the examined data set.\\

Using the aforementioned formalism and $U=6$ 2-dimensional 
examined functions from Fig.~5-8 (symbols) and 1-dimensional
functions from Fig.~9,10, 
the unknown spectral parameters $\Phi(A),E_k(A),
\gamma_2$ were derived by the minimization of $\chi^2$ function (3) at 
$\gamma_1=2.65$ and the degree of freedom $\sum_{1}^{6}{V_u}\simeq 350$.\\
The values of spectral parameters (2) obtained by the solution
of the parameterized equation (1) are presented in Table~1
at the QGSJET and SIBYLL interaction models. 
\begin{table}
\caption{\label{tab:table1}
Parameters of primary energy spectra (2) 
at 1,2-dimensional analysis of EAS data. Scale factors $\Phi_A$
and particle's rigidity $E_R$
have units of $(m^2\cdot sec\cdot ster\cdot TeV)^{-1}$
and ($TV$) respectively.}
\begin{ruledtabular}
\begin{tabular}{lcc}
Parameters & SIBYLL&QGSJET\\
\hline
$\Phi_  H$ & $0.081\pm0.004$ & $0.164\pm0.004$\\
$\Phi_{He}$& $0.072\pm0.008$ & $0.005\pm0.008$\\
$\Phi_O   $& $0.028\pm0.008$ & $0.005\pm0.006$\\
$\Phi_{Fe}$& $0.028\pm0.003$ & $0.018\pm0.003$\\
$E_R      $& $2560\pm200$    & $3400\pm150$   \\
$\footnote{Parameter was fixed.}\gamma_1$ & $2.65$ & $2.65$\\
$\gamma_2$ & $3.21\pm0.04$   & $3.10\pm0.03$\\
$\chi^2$  & $2.5$ &$2.6$\\
\end{tabular}
\end{ruledtabular}
\end{table}
The derived primary energy spectra for $p, He, O, Fe$ nuclei 
are shown in Fig.~11 (shaded areas) in comparison with the  
KASCADE data (symbols) from \cite{KASCADE04}.\\
The expected spectra conforming the examined data set
according to the solutions above are shown in Fig.~5-10 (lines
and shaded area) for the QGSJET and SIBYLL interaction models.
It is necessary to note, that the obtained results
in the framework of the SIBYLL interaction model are more consistent and
slightly dependent on a number of examined functions.
%===========================================
\subsection{4-Dimensional approach}
The combination of 1,2-dimensional approximations of EAS data above does not take 
into account
all the information about primary energy spectra 
folded in the detected EAS data. 
In general, the EAS inverse problem can be formulated
in the multidimensional space of EAS parameters.
In case of the 4-parametric ($N_{ch},N_{\mu},s,\theta$) analysis,
the expression (1) is written as:
\begin{eqnarray}
\frac{\Delta F}{\Delta\tilde{N}_{ch}\Delta\tilde{N}_{\mu}\Delta 
\tilde{s}\Delta\Omega}=
&&\frac{1}{C}
\sum_{A}\int_{E} \frac{dI_A}{dE}
\int_{Q}\int_{D} 
{\cal G}_A(E,\theta)
\nonumber \\
&&\times
{\cal R}(\theta)
dEdDdN_{ch}dN_{\mu}ds\;,
\end{eqnarray}
where ${\cal G}_A(E,\theta)\equiv\partial^3 W_A(E,\theta)/
\partial N_{ch}\partial N_{\mu}\partial s$, 
are the multidimensional differential EAS spectra
at given $A,E,\theta$ parameters of the primary nucleus,
${\cal R}(\theta)\equiv
\partial^3 R(\theta)/\partial\tilde{N}_{ch}\partial\tilde{N}_{\mu}\partial
\tilde{s}$
are the error functions of the experiment. 
The parameters with a tilde symbol are the reconstructed values 
of corresponding EAS parameters.\\
Evidently, the amount of 
information about primary energy spectra contained
in the detected multidimensional spectrum 
$\Delta F$ 
is always greater than the cumulative amount of information contained 
in the 1,2-dimensional spectra $\Delta F_u/\Delta\tilde{q}_u$, 
$\tilde{q}_u\equiv\tilde{N}_{ch},\tilde{N}_{\mu},\tilde{s}$ 
of the expression (1). The difference is determined
by the inter-correlations of EAS parameters that are taken
into account in the expression (4).\\
On the basis of the EAS data set of the GAMMA experiment, 
the simulated EAS database (section III) and parameterization (2), 
the equations (4) were resolved by the $\chi^2$-minimization method. 
The computations were performed at the following bin dimensions: 
$\Delta\ln{N_{ch}}=0.15$, 
$\Delta\ln{N_{\mu}}=0.25$,
$\Delta\sec{\theta}=0.05$
and $\Delta s=0.15$ on the left and right hand side of $s^*=0.85$ and
$\Delta s=0.3$ in other cases. 
The total number of the degree of freedom at 4-dimensional 
$\chi^2$-minimization was equal to $1560$.\\ 
The values of spectral parameters (2) obtained by the solution
of the parameterized equation (4) are presented in Table~2
at the QGSJET and SIBYLL interaction models. 
\begin{table}
\caption{\label{tab:table2}
Parameters of primary energy spectra (2) 
at 4-D analysis of EAS data. Scale factors $\Phi_A$
and particle's rigidity $E_R$
have units of $(m^2\cdot sec\cdot ster\cdot TeV)^{-1}$
and ($TV$) respectively.}
\begin{ruledtabular}
\begin{tabular}{lcc}
Parameters & SIBYLL&QGSJET\\
\hline
$\Phi_  H $ & $0.089\pm0.003$ & $0.14\pm0.004 $\\
$\Phi_{He}$ & $0.053\pm0.005$ & $0.034\pm0.004$\\
$\Phi_O   $ & $0.049\pm0.004$ & $0.016\pm0.003$\\
$\Phi_{Fe}$ & $0.029\pm0.003$ & $0.015\pm0.002$\\
$E_R      $ & $3000\pm300   $ & $3300\pm200   $\\
$\gamma_2 $ & $3.16\pm0.08  $ & $3.10\pm0.03$  \\
$\chi^2   $ & $1.2          $ & $1.1$          \\
\end{tabular}
\end{ruledtabular}
\end{table}
As it is seen from Fig.~11 and Tables~1,2, the derived expected primary 
energy spectra significantly depend on interaction model.
The expected abundance of primary nuclei at energy $E\sim10^3$ TeV
in the framework of SIBYLL model agrees well with corresponding 
extrapolations of the balloon and satellite data \cite{Wiebel}, 
whereas the predictions according to the QGSJET model point out
to a predominantly proton primary composition in the $10^3\div10^5$
TeV energy range.
%==================================================
\section{Event-by-event analysis}
The mountain location of the GAMMA experiment and the agreements of observed
and simulated data in the measurement range $5\cdot10^5\leq 
N_{ch}<10^7$ (Fig.~4-10) allowed, apart from above, 
to obtain the all-particle energy spectra with high reliability . 
The method is based on an event-by-event 
evaluation of primary energy using the reconstructed parameters 
$\tilde{N}_{ch},\tilde{N}_{\mu},\tilde{s},\theta$ of detected EAS. 
Such possibilities have been studying for a long time in
different papers \cite{Capdevil, GAMMA02, Ter-Antonyan6} 
and the main difficulty
was to obtain an unbiased energy estimation at an existent 
abundance of the primary nuclei taking into account the fluctuations
of shower development and detector response.\\
Using the simulated database, $J=1.5\cdot10^4$ 
EAS events were taken for each of $k=1,\dots,4$ kinds ($H,He,O,Fe$) of primary 
nuclei
and each interaction model (SIBYLL, QGSJET). 
The reconstructed $\tilde{N}_{ch},\tilde{N}_{\mu},\tilde{s}$ shower parameters,
known zenith angle $\theta$ and primary energy $E_0$ were used at 
minimization 
\begin{equation}
\chi^2(a_1,\dots,a_p)=
\frac{1}{4J}
\sum_{k=1}^{4}
{\sum_{j=1}^{J}
{\frac{(\ln{E_{1,k,j}}-\ln{E_{0,k,j}})^2}{\sigma^2_E}}}
\end{equation}
where $E_1=f(a_1,\dots,a_p |\tilde{N}_{ch},\tilde{N}_{\mu},\tilde{s},\theta)$ 
is the investigated
parametric function with $a_1,\dots,a_p$ parameters,
$\sigma_E$ is expected accuracy of the $E_1$ evaluated energy.  
\begin{figure}[h]
%\begin{center}
%\epsfig{file=fig13.eps,width=6cm,height=6cm}
\includegraphics[width=7cm,height=7cm]{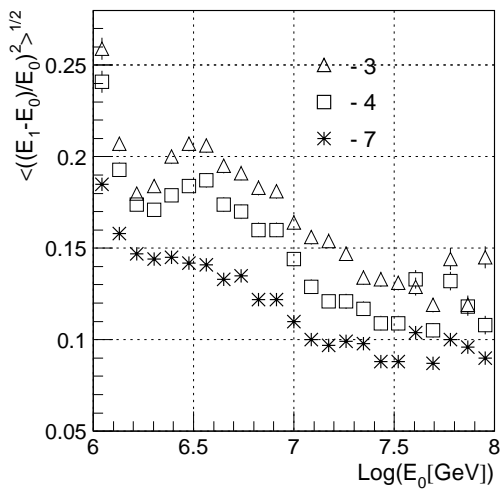}
%\end{center}
\caption{Accuracy (RMSD) of the primary energy evaluations at different
number of approximation parameters.}
\end{figure}
The best estimations were achieved at 7-parametric ($p=7$)
fit:
\begin{equation}
\ln{E_1}=a_1x+\frac{a_2s}{c}+a_3+a_4c
+a_5e^s+ \frac{a_6}{(x-a_7y)},
\end{equation}
where $x=\ln{\tilde{N}_{ch}}$, $y=\ln{\tilde{N}_{\mu}(R<50m)}$,
$c=\cos{\theta}$ and energy $E_1$ has units of GeV.
The values of the 
$a_1,\dots,a_7$ parameters for both interaction models and the corresponding 
$\chi^2$ obtained from (5) at $\sigma_E=0.15$ are displayed in Table~3.
\begin{table}
\caption{\label{tab:table3}
Approximation parameters $a_1,\dots,a_7$ of primary energy evaluation
(6) and corresponding $\chi^2$ obtained from (6) at the SIBYLL and QGSJET
interaction models.}
\begin{ruledtabular}
\begin{tabular}{lcccccccr}
Model&$a_1$&$a_2$&$a_3$&$a_4$&$a_5$&$a_6$&$a_7$&$\chi^2$\\
\hline
SIBYLL&1.03&3.98&-4.3&2.01&-1.2&11.8&0.94&0.85\\
QGSJET&1.03&4.38&-4.6&2.35&-1.3&11.5&0.96&0.94\\
\end{tabular}
\end{ruledtabular}
\end{table}
The root mean square deviations of the energy estimation 
by 7-parametric fit (4) in the framework of the SIBYLL model is 
shown in Fig.~12. The corresponding results 
at three (only $x,\tilde{s}$ variables) and 4-parametric 
($x,\tilde{s},\cos{\theta}$)
fit are shown in Fig.~12 as well.\\
\begin{figure}[h]
%\begin{center}
%\epsfig{file=fig14.eps,width=8cm,height=8cm}
\includegraphics[width=7cm,height=7cm]{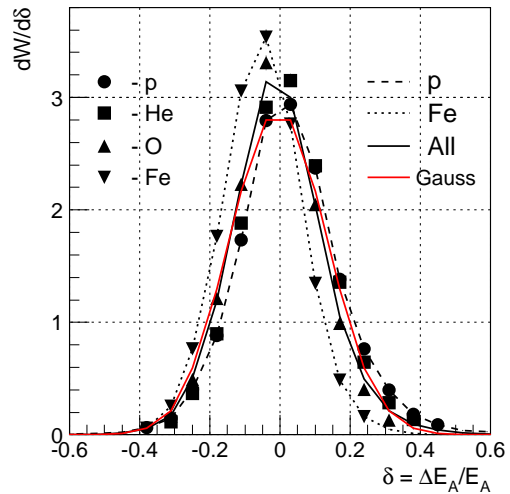}
%\end{center}
\caption{Distribution of errors of the primary energy estimation 
by event-by-event 7-parametric fit at different primary nuclei.}
\end{figure}
The obtained error distributions estimating primary energy
by 7-parametric approximation (6), are shown in Fig.~13 for $H,He,O,Fe$ 
nuclei. The red line corresponds to the Gaussian distribution at the
same parameters as cumulative distribution (black solid line).
\\
The all-particle energy spectrum derived on the basis
of the  GAMMA 2002-2004 EAS data set and fit (6), 
at the QGSJET (filled red square symbols) and SIBYLL (filled blue circle 
symbols)
interaction models are shown in Fig.~14.\\
Notice, that the energy spectrum obtained by event-by-event method 
claims additional corrections, because the errors $\sigma_E=\sigma(\Delta E/E)$ 
and power-law energy spectra  ($\sim E^{-\gamma}$) lead to
an overestimation of the spectrum $\eta=\exp{(((\gamma-1)\sigma_E)^2/2)}$ 
times.
Moreover, the inevitable biases of energy estimations
$\epsilon(A)=<E_1/E_0>$ depend on primary nuclei and shift the corresponding 
energy
spectra  $\beta(A)=\epsilon^{\gamma-1}$ times. The spectral shift due to 
$\beta(A)\neq1$
impossible to take into account without information about abundance 
of primary nuclei.\\
The observed biases of 7-parametric fit (6) are distributed from 
$\epsilon(p)\simeq1.02\%$ up to $\epsilon(Fe)\simeq0.96\%$ (Fig.~13) and 
here are neglected. In the results shown in Fig.~14, the corrections of 
$\eta(E)$ are taken into account using the expected accuracies from Fig.~12.\\
The solid (red) and dashed (blue) lines in Fig.14 represent the all-particle
primary energy spectra obtained on the basis of GAMMA data set by the solution 
of parametrized EAS inverse  problem in the framework of the SIBYLL and 
QGSJET models respectively.
\begin{figure}[h]
%\begin{center}
\includegraphics[width=8cm,height=8cm]{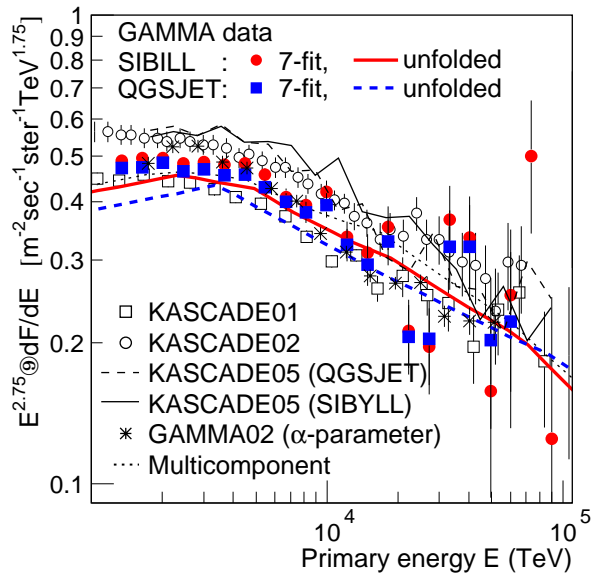}
%\end{center}
\caption{All particle primary energy spectra obtained by event-by-event
7-parametric analysis (filled symbols) and EAS inverse problem solutions 
(solid and dashed lines) on the basis of GAMMA 2002-2004 database.}
\end{figure}
The event-by-event analysis of the GAMMA data at the QGSJET interaction model
using $\alpha$-parametric method \cite{GAMMA02} also shown in Fig.~14 
(asterisk symbols). The dotted line in Fig.~14 represents the parametrized 
solutions of the EAS inverse problem for the KASCADE EAS data at 
rigidity-dependent steepening primary energy spectra \cite{Ter-Antonyan5}. 
The results of KASCADE02 in Fig.~14 obtained by the non-parametric
event-by-event analysis was taken from review \cite{Swordy}. 
The KASCADE01,05 data obtained by the iterative method \cite{Gold}
of unfolding 
of the EAS inverse problem were taken from \cite{KASCADE01,KASCADE05}
respectively.
%==================================================
\section{Conclusion} 
The self-consistency of results (Fig.~2-10) obtained by GAMMA experiment 
at least up to $N_{ch}\simeq10^7$ and corresponding predictions in the 
framework of hypothesis of the rigidity-dependent steepening primary energy 
spectra and validity of the SIBYLL or QGSJET interaction models 
point towards:
\begin{itemize}
\item The anomalous behavior of the EAS muon spectra (overestimation, 
Fig.~4b,8,10)
and EAS age parameter at EAS size $N_{ch}>10^7$. The same behavior 
of the EAS age parameter had been observed also in \cite{Norikura,MAKET}. 
\item The obtained abundances and energy spectra of primary $p$, $He$, $O$, 
$Fe$ nuclei depend on interaction models. 
The SIBYLL interaction model is more preferable in terms of 
the extrapolation of the derived expected primary spectra (Fig.~11) to the 
energy range of the direct measurements.
\item The rigidity-dependent steepening energy spectra of primary nuclei
describe the EAS data of the GAMMA experiment at least up to $N_{ch}\simeq10^7$
with average accuracy $<10\%$ at particle's magnetic rigidity 
$E_R\simeq2400\div3000$ TV (SIBYLL) and $E_R\simeq3300\pm200$ TV (QGSJET).
\item The 4-dimensional approach at the EAS inverse problem solution is more 
preferable in terms of the stability and accuracies of solutions.  
\item The obtained all-particle energy spectra slightly depend on
interaction model and are practically the same at both the event-by-event 
reconstruction method and the EAS inverse approach.
\end{itemize}
The obtained energy spectra of primary nuclei ($A\equiv p,He,O,Fe$) in 
the energy range $10^6<E_A<5\cdot10^7$ GeV disagree (see Fig.~11) with 
the same KASCADE data obtained by the iterative method \cite{Gold}. 
However, the 
discrepancies of all-particle energy spectra (see Fig.~14)
obtained by the GAMMA and KASCADE 
experiments are sufficiently small ($\sim20\% $). 
%==================================================
\section*{Acknowledgments}
We are grateful to all of our colleagues at the Moscow Lebedev 
Institute and the Yerevan Physics Institute who took part 
in the development and exploitation of the GAMMA array.\\
This work has been partly supported by research grant No 1465 
from the Armenian government, NFSAT bilateral 
U.S.-Armenian grant AS084-02/CRDF 12036, 
by the CRDF grant AR-P2-2580-YE-04  and the "Hayastan" 
All-Armenian Fund in France.
%==================================================

\end{document}